\documentclass[11pt]{article}

\usepackage{hyperref}

\usepackage{amsmath}
\usepackage{amsthm,amssymb}
\usepackage{graphicx}
\usepackage{algorithm}
\usepackage{algorithmic}
\usepackage{xspace}

\newcommand{\eps}{\epsilon}

\newcommand{\E}{\mathbb{E}}

\newcommand{\opt}{\textsc{opt}}
\newcommand{\alg}{\textsc{alg}}

\newcommand{\cD}{\mathcal{D}}

\newcommand{\cR}{\mathcal{R}}
\newcommand{\cA}{\mathcal{A}}
\newcommand{\cH}{\mathcal{H}}
\newcommand{\cP}{\mathcal{P}}
\newcommand{\cE}{\mathcal{E}}

\newtheorem{theorem}{Theorem}
\newtheorem{lemma}{Lemma}[section]
\newtheorem{example}{Example}[section]
\newtheorem{define}{Definition}[section]
\newtheorem{corollary}[theorem]{Corollary}

\newcommand{\poa}{{\sc Poa}\xspace}
\newcommand{\pico}{{\sc Pico}\xspace}

\begin{document}

\newcommand\relatedversion{}
\renewcommand\relatedversion{\thanks{The full version of the paper can be accessed at \protect\url{https://arxiv.org/abs/1902.09310}}} %

\title{\Large Online Distributed Queue Length Estimation}

\author{
  Aditya Bhaskara\thanks{
  School of Computing 
  University of Utah
  (\texttt{bhaskaraaditya@gmail.com}).
  Supported by NSF awards CCF-2008688 and CCF-2047288. 
  } \and
  Sreenivas Gollapudi \thanks{
  Google Research, 
  Mountain View, CA (\texttt{sgollapu@google.com}, \texttt{kostaskollias@google.com}). 
  }  \and 
  Sungjin Im \thanks{University of California, Santa Cruz, CA (\texttt{sim9@ucsc.edu}). 
  Supported in part by NSF grants  CCF-1844939, CCF-2121745, and CCF-2423106, and ONR grant N00014-22-1-2701.} 
  \and
  Kostas Kollias \footnotemark[2]  \and
  Kamesh Munagala\thanks{Computer Science Department, Duke University (\texttt{munagala@duke.edu}). Supported by NSF grant CCF-2113798 and IIS-2402823.}  
}

\date{}

\maketitle

\begin{abstract} Queue length monitoring is a commonly arising problem in numerous applications such as queue management systems, scheduling, and traffic monitoring. Motivated by such applications, we formulate a queue monitoring problem, where there is a FIFO queue with arbitrary arrivals and departures, and a server needs to monitor the length of a queue by using \emph{decentralized} pings from packets in the queue. Packets can send pings informing the server about the number of packets \emph{ahead} of them in the queue. Via novel online policies and lower bounds, we tightly characterize the trade-off between the number of pings sent and the accuracy of the server's real time estimates. 
Our work studies the trade-off under various arrival and departure processes, including  constant-rate, Poisson, and adversarial processes.
\end{abstract}

\section{Introduction} \label{sec:intro}

FIFO queues are a ubiquitous modeling tool in various applications such as job scheduling, packet routing, buffering, service request handling, etc. In this paper, we consider FIFO queues that consist of packets representing \emph{distributed agents}. This is a natural way to model scenarios like cars moving on a congested road segment or waiting at a traffic light, customers lining up for their turn at a service counter, and so on. We aim to study the problem of monitoring the status of such a queue, in particular, keeping track of the length of the queue as the agents (who we will call \emph{packets}) enter and leave the system. We consider the setting where the central server aims to maintain a real time estimate of the queue length, using \emph{pings} received from the packets. Our goal is to understand the following questions: What are the tradeoffs between the number of pings sent to the server (i.e., total communication) and the accuracy of the estimates? What \emph{policy} should individual packets follow in order to achieve the desired bounds?

\smallskip \noindent {\bf Motivation.} Queue length is commonly used to estimate a client's waiting time in the line for service. Tracking the length is highly important for various reasons. It can be used by the service provider for effective resource management and also by the clients to choose the best among multiple queues available to them. If the queue length can be monitored by a dedicated sensor, the problem is trivial. However, it becomes a significant challenge when the monitoring relies on packets that cannot observe the global queue length and are controlled by distributed and independent agents. 

This situation arises particularly in traffic monitoring~\cite{SGS+17,G07,MW11,YYL04,FRV08,TBO12}. Monitoring the congestion on various road segments in real time is a central component of routing systems (whose goal is to find the best route for a user). The monitoring server  (which we will simply refer to as the server) receives information about the ``congestion'' on a segment from cars on the segment. These cars, in turn, use \emph{local} information such as their speed, location, etc. to determine what to communicate to the server and when to do so. The natural measure of congestion  is the total number of cars on the segment. 

We need to consider both the \emph{pinging policy} followed by the individual cars (or devices located in cars), as well as the server-side \emph{reconstruction of the congestion}. We have two main design goals. The first one is the ping frequency. Pinging very frequently may allow the server to better estimate congestion, but it also drains the battery life on GPS devices, which are typically cellphones, and may compromise user privacy. The second goal is to have decentralized one-way pings. In other words, each device acts in isolation without coordinating pings with either the server or with other devices. This is motivated by issues of trust and privacy, as well as the cost of communication. 

For designing policies achieving the goals above, we must first consider what information is available to the distributed devices. In the traffic setting, a device in a car has very little global information about the full road segment. It has access to the current position and speed of the car, but these are quantities that are only affected by the cars {\em ahead}, not cars behind. Detailed discussion on how the number of cars ahead can be inferred can be found in the related work for the Car Following Model (Section~\ref{sec:related}). Likewise, in the setting of people waiting at a service line, a person may know how many people are ahead, but not the total length of the line. Motivated by these goals and observations, we formulate the following {\em queue monitoring} problem.

\smallskip \noindent {\bf Queue Monitoring Problem.} Suppose we have a FIFO queue to which packets arrive and depart at discrete time steps according to arbitrary arrival and departure processes. A {\em server} needs to continuously monitor the queue length, which we use as a proxy for congestion in the queue. At any time instant, any packet can send a {\em ping} to the server notifying it about the number of packets {\em ahead} of it in the queue (along with any other information it sees fit). The system is decentralized and communication is one-way, so the server cannot send information to any specific packet, nor can the packets communicate with each other. The goal in this setting is to design a {\em pinging policy} for the packets that uses only locally available information, and a {\em recovery procedure} for the server that lets it use the pings received so far to maintain an estimate of the queue length at any point of time.

We assume that at any time, a packet in the queue knows how many packets are ahead of it, but has no further information. Thus, a pinging decision is made based on this information collected so far. On the packet side, the goal is to minimize the rate of pings. On the server side for recovery, our error measure is the average (over time) of the difference between the prediction and the true number of packets in the queue, i.e., the difference between the predicted and the true queue length.

\smallskip
The queue monitoring problem is one instance of a more general class of problems where we have distributed agents that each use myopic and local information to send information to a server that aims to keep track of a global state. Our algorithms and lower bounds can thus provide templates to reason about other such scenarios.

\subsection{Our Results}
Our main contribution is the design and analysis of simple yet novel pinging policies for the queue monitoring problem. Our policies adaptively regulate their ping probability based on locally available information, and we show that the server can indeed accurately reconstruct the queue length. In our policies, a packet (at any time) sends a ping with a probability that is a function $g(h,w)$, where $h$ is the number of packets \emph{ahead} of it, which we term its height, and $w$ is its waiting time, the time elapsed since it entered the queue. We further establish the optimality of the functional forms $g(h,w)$ that appear in our algorithms.

We develop two types of pinging policies: the first is termed \poa (Section~\ref{sec:poa}), wherein a packet only pings upon arrival into the queue, and the second is termed \pico (Section~\ref{sec:pico}) where the packet pings continuously, at an adapting rate. The former has the advantage that a packet sends at most one ping, and does not need to keep track of any other state until it exits the queue. However, \poa is insufficient for worst case  arrival and departure processes, and we need continuous pinging. The details of our results are as follows.

\smallskip
\noindent {\bf Ping on Arrival.} As described above, in \poa policies, packets send pings only on arrival. Due to the limitation on the information available to any packet, (distributed) \poa policies have a very simple structure: if a packet arrives and observes the queue to be of height $h$, it sends a ping with probability $f(h)$. Note that the observed height is equal to the queue length for \poa policies. 
We analyze \poa policies in Section~\ref{sec:poa}. We show that despite seeming restrictive, \poa policies can yield non-trivial results. When departures occur at a constant rate while arrivals are adversarial, the optimal pinging policy belongs to this class. We then develop a \poa policy that achieves reconstruction error $\epsilon$ (defined formally in Section~\ref{sec:model}) times the average queue length, using $f(h) = \tilde{O}\left(\min\left(1,\frac{1}{\epsilon h}\right)\right)$, where the $\tilde{O}$ hides terms depending on $\ln \frac{1}{\epsilon}$.  
We prove the optimality by showing in Section~\ref{sec:lb_dep} that any monotone pinging function $f(h)$ needs to be at least this value for \poa policies that achieve $\epsilon$ reconstruction error. We further show in Appendix~\ref{sec:ub2} that our upper bound for the \poa policy with constant-rate departures easily implies the same guarantee when departures follow a Poisson process ($G/M/1$ queueing model~\cite{grimmett2001}), while arrivals can be arbitrary.  

In Appendix~\ref{sec:unif_arrival}, we consider the `opposite' case, where arrivals are constant-rate and the departure process is adversarial, and extend this to Poisson arrivals in Appendix~\ref{sec:ub4}. 
For all these settings, we not only derive similar guarantees for the \poa policy, but also show matching lower bounds (in Section~\ref{sec:lb_dep}), thus justifying the functional forms of our \poa policies. 

\smallskip
 \noindent {\bf Ping Continuously.} The second type of policies we consider (in Section~\ref{sec:pico}) allow packets to ping at any time in which they are in the queue. We show that such policies are needed (and \poa policies are insufficient) when both departures and arrivals either follow more bursty stochastic processes than Poisson, or are entirely adversarial. In such a policy, that we term \pico, each packet pings every time step with probability $g(h,w)$, where $h$ is its height and $w$ is its waiting time (the number of time steps elapsed) since its arrival. We show that for any adversarial arrival \emph{and} departure processes, when $g(h,w) = O\left( \frac{\ln (1/\epsilon)}{\epsilon^2} \frac{1}{hw}\right)$, then the reconstruction error in queue length is at most $\epsilon$. We finally show that the number of pings generated by this policy for a packet arriving at height $h$ is $O\left(\frac{\log h}{\eps}\right)$ factor more than that of the \poa policy when either arrivals or departures occur at a constant rate, showing there is not much room for improvement in the functional form.

\smallskip
\noindent {\bf Beyond Monitoring a Single Queue.}
Although we focus on monitoring a single queue, we remark that this is not too restrictive. We note that our results yield guarantees for networks of queues via known results. In this case, the reconstruction error is the sum of the errors in the queue heights, and we can bound this error by $\epsilon$ times the sum of the queue lengths. From our results on the \poa policy for Poisson departures (see Appendix~\ref{sec:poisson}), the same guarantee holds for a Jackson network of queues where all departure processes are Poisson~\cite{grimmett2001}. Similarly, the \pico policy in Section~\ref{sec:pico} extends to a network of queues with arbitrary arrival and departure processes.

\subsection{
 Algorithmic Intuition and Technical Contributions}  To build intuition, first assume that packets arrive, but do not depart. In this case it is natural to get the queue length from the latest arriving packet whenever the queue length increases by a certain factor. This can be simulated by the \poa policy that pings with a probability inverse to the height, $h$. However, it is unclear if this policy would continue to work when packets can depart the system, since the arrival and departure processes can interleave in a complicated way. We show that \poa policies do work when one of the arrival or departure processes is somewhat predictable (constant-rate or Poisson), while the other process is adversarial.

For further intuition behind our \poa policy, consider the setting where packets depart at a constant rate or according to a Poisson distribution. Since job departure is predictable, the only possible scenario for the server to be significantly outdated with the current queue height occurs when a large number of packets arrive within a short period of time. In this case, \poa is designed so that at least one packet sends a ping to the server with a good probability before the height changes by a factor of more than $1 + \eps$.

 However, the analysis is challenging. For each packet $i$, we measure the lag $\delta_i$ of it as the expected time it takes for the server to become aware of it by receiving a ping from it or its subsequent packet (packet $i$ is counted in the height of a subsequent packet upon arrival if $i$ is still present). We carefully charge $\E \sum_i \delta_i$ to an $\eps$ fraction of the total delay. Here, we combine ideas from competitive analysis, queueing theory, and random processes.

When the arrival and departure processes are both adversarial, we need an entirely different policy. With a \poa policy, if a large number of packets depart instantly and few new packets arrive, the server will remain clueless of this change. We argue that we have to allow packets to ping intermittently.   The basic intuition behind our second algorithm \pico is to view each packet $i$'s height $h$ at each time $t$ as a point $(t,h)$. Suppose we are only interested in estimating the total delay  of packets over all times. This delay is just the number of all points. We can now let this point $(t,h)$ add weight $h \cdot w$ to the delay estimate with a probability $\frac{1}{hw}$, where $w$ is the packet's waiting time. This will give an unbiased estimator of total delay.

However, the goal of the server is to accurately interpolate the true delay, meaning that we should estimate the number of packets present in the queue at \emph{each} time. To do this, we consider the rectangle associated with the point, which has width $w$ and height $h$. That is, this rectangle $R$ has four vertices: $(t - w, h)$, $(t, h)$, $(t - w, 0)$, and $(t, 0)$. Note that this rectangle lies below the height profile curve as a function of time because the number of packets ahead of packet $i$ can only decrese in time. Thus, the union of such rectangles lies entirely below the profile curve. Further, the \pico ping rate is designed so that this union is close to the profile curve.

Then, the server projects each rectangle encoded by a ping forward in time and increases its height---the magnified rectangle has four vertices: $(t - w, h)$, $(t + 3\epsilon w, (1 + 3\epsilon)h)$, $(t - w, 0)$, and $(t + 3\epsilon w, 0)$. The server uses the union of these magnified rectangles as its estimate. The analysis shows that this union almost completely covers the profile curve while being only slightly larger than the union of the original rectangles.

\subsection{Related Work}
    \label{sec:related}
\noindent {\bf Online Buffer  Management.}  There are several online models for queue management that are motivated by networking applications; this includes dynamic TCP acknowledgment~\cite{stories,BN}, buffer reordering~\cite{buffer,ImMoseley}, and packet scheduling in switches~\cite{Andelman}. However, in the TCP acknowledgment problem (and more generally buffer management), the entity sending acks is centralized and has complete information about arrivals so far. In contrast to this model, our setting is decentralized and each pinging vehicle (packet) has incomplete information about both the total delay and which other vehicles are simultaneously sending pings. Therefore, the decentralized aspect in our problem makes the typical benchmark of a centralized offline optimum used in competitive analysis too strong. Part of our innovation therefore is in coming up with the right analytic framework to show the optimality of our policies.

\smallskip \noindent {\bf Stochastic Queue  Management.} Similarly, several recent works in queueing theory~\cite{JIQ,scalable} have considered the problem of stochastic load balancing of multiple queues when the assigner gets limited feedback from the queues about their load. The focus of this work is to derive policies for sending feedback under which the scheduling policy is {\em stable}. In contrast with this work, our motivation comes from routing applications, where giving the end-user accurate feedback about delay on their route is crucial. This makes our goal different: we seek to {\em monitor} the length (or height) of the queue via feedback, and this is interesting even in single-queue setting where load-balancing is a trivial problem. Further, at a technical level, the performance of our policies crucially depend on the {\em nature} of the arrival and departure processes, and merely assuming ergodicity is  insufficient. (See Section~\ref{sec:lb_dep} for lower bounds for general ergodic processes.) 

\smallskip \noindent {\bf Energy-efficient Functional Monitoring.}  The general problem of {\em distributed functional monitoring}~\cite{CormodeMY08,WoodruffZ12,TopKStream} has received considerable attention in the context of energy efficiency in sensor networks. In this setting, a set of distributed low-power sensor nodes receive an arbitrary sequence of items over time. These nodes need to communicate as few bits of information as possible to the server (to save energy), so that the server can maintain a good running approximation to some statistic over all the items, such as frequency moments, heavy hitters, etc. This model generalizes the classical streaming model~\cite{AMS}. In the distributed monitoring model, nearly tight bounds on the trade-off between communication and approximation are known for several basic statistics, even when item insertions are adversarial. However, when deletions are allowed, the problem becomes hard. Our problem can be viewed as functional monitoring, where the function is the delay in the queue. The key difference is that the vehicles are sensors, and they arrive and depart the system instead of being fixed. Further, these observations are correlated with each other and across time via traffic dynamics. This makes our problem technically different.

 \smallskip \noindent {\bf Delay Estimation for Traffic Networks.} We briefly review relevant work in traffic monitoring. Existing work in this area largely focuses on using sparse and noisy GPS traces to accurately estimate historical traffic, real-time traffic or future traffic~\cite{HHA+09,HHA+10,EJT+15}.  Solutions for estimating route-specific ETAs have been proposed using ML models based on physical world features such as weather conditions, vehicle type, current network congestion, etc.~\cite{YZX+11,ZKJ18}. The work most closely related to ours is~\cite{ZZW+19}. This work assume a uniformly random subset of the vehicles are ``probe'' vehicles that continuously update the server about queue lengths ahead of them ({\em i.e.}, their stop positions at intersections). They develop various statistical estimators for the actual queue lengths based on the positions of these probe vehicles.    
In contrast, we assume vehicles cannot continuously update the server, and this leads to the natural question of adaptively minimizing the number of probes needed to generate good estimates of the  delays.  

\smallskip \noindent {\bf Car Following Model.}
Since we assume that a packet can infer the number of packets ahead of it in our model, we explicitly discuss why this is a reasonable assumption in the contexts of traffic congestion. A vehicle can infer the number of vehicles (or congestion) ahead of it as follows. If the vehicle is equipped with technology that enables it to calculate following distance (say via radar) and its exact location, it can set the local estimate of congestion as $\ell/\delta$, where $\ell$ is the distance to the head of the segment, and $\delta$ is the following distance.  If the quantity $\delta$ is not directly available, this can be inferred with a simple and well-known model called the {\em simplified car following} model of Newell {\em et al.}~\cite{newell2002simplified}. In this model, the segment is assumed to have a uniform speed $v$, and given this speed, the vehicles have an average ``following distance" $s$ given by $s = d + \tau v$, where the parameters $d$ and $\tau$  can be learnt from historical data. These two parameters are average intrinsic properties of drivers, for instance $\tau$ represents the time needed to safely brake if the previous car suddenly stops. Suppose that a vehicle knows its speed $v$ and its location $\ell$ relative to the head of the segment, then it can estimate  the number of vehicles ahead of it as $ h = \frac{\ell}{d + \tau v}.$   Alternatively, the mapping from speed $v$ to $h$ can be learnt (say per segment) from historical data and the resulting ML model can be used by the vehicle. Note that since a car only knows its location and speed (and not about future arrivals into the segment), it can only infer the number of vehicles ahead of it in the segment, motivating our assumption. 

\section{The Queue Monitoring Problem}
\label{sec:model}
We consider a FIFO queue into which  packets arrive and depart. Packets arrive at the tail of the queue and depart at the head. Let $a_i$ denote the arrival time of packet $i$. Departures can be viewed as follows: There is a stream of ``departure tokens''; when a departure token is generated, if there is at least one packet in the queue, the packet at the head of the queue departs. Let $d_i$ denote the departure time of packet $i$. The {\em delay} experienced by this packet (i.e., the time spent in the queue) is thus $d_i - a_i$. Let $\cA$ denote the set of arrival times $\{a_1, a_2, \ldots\}$ and $\cD$ denote the set of departure times $\{d_1, d_2, \ldots, \}$. We will assume throughout that time is discrete, and thus $a_i, d_i$ are all integers. There is a horizon of $T$ steps at the end of which all packets have departed.

At time $t$, we denote the number of packets in the queue as the ``height'' $h_t$. We have the following equality between the time-averaged delay and average height (which we call $\opt$).
\begin{equation}
    \label{eq:optt}
\opt := \frac{1}{T} \sum_{i \in S} (d_i - a_i)  =  \frac{1}{T} \sum_{t=0}^T h_t. 
\end{equation}
This holds because the sum on the LHS measures the number of time steps packet $i$ is in the queue, and at each such step, it contributes a value of $1$ to the height, which is the RHS sum. 

 \smallskip \noindent {\bf Queue monitoring policy.}  Packets in the queue send pings to a central monitoring server (henceforth called the server) using a pinging algorithm $\cP$, and the goal of the server is to keep track of $h_t$ at every time step $t$, using an estimation algorithm $\cE$. Thus we define queue monitoring policy as the pair $(\cP, \cE)$. We assume that the algorithm $\cP$ is fully decentralized, and there is no communication between packets.

Next, we have the main twist that we motivated earlier: each packet in the queue only knows the number of packets ahead of it in the queue. The packet does not know of the existence of later arrivals into the queue, and in particular, it need not know the current height $h_t$ of the queue.  
We let $h_{it}$ denote the ``position'' of packet $i$ in the queue (i.e., number of packets ahead of it including itself) at time $t$, for $t \in [a_i, d_i]$. Our key assumption is that the pinging algorithm $\cP$ (which decides the probability of sending a ping and the information to include in each ping) depends only on $i$'s {\em information set} at time $t$, which is the set of $h_{it'}$ values at all times $a_i \le t' \le t$.\footnote{While this might sound restrictive, most natural policies, e.g., pinging when the `speed' changes can be captured here (e.g., by looking at the rate of change of the $h$ values).}

On the server side, the estimation algorithm $\cE$ must use the pings received from the packets to  
maintain a real-time estimate of the queue height. We denote the estimate at time $t$ by $e_t$. Note that the server does not have direct access to the queue and estimates only using the pings. 

We denote by $\alg$ the time-average error in the server's estimate of queue height. Since the algorithms $\cP$ and $\cE$ can be randomized, we formally define it as an expected value:
\begin{equation}
    \label{eq:algt}
\alg = \frac{1}{T} \E\left[\sum_{t=0}^T | h_t - e_t |  \right].
\end{equation}

 \smallskip \noindent {\bf Design goals.}  Our aim is to design a pinging policy such that (i) the total number of pings is small, and (ii) the average error $\alg$ in the server's estimate of queue height is small. Formally, let $\epsilon \in (0,1)$ be a small constant that is given as an error parameter. Our goal is to design a pinging policy with $\epsilon$ reconstruction error, which we define as: 
\begin{equation}
    \label{eq:eps}
    \alg \le \epsilon \cdot \opt + c
\end{equation} 
for some absolute constant $c$ (recall that $\opt$ was defined in~\eqref{eq:optt}). For this guarantee to be meaningful, we assume that either the arrival or departure process is sufficiently variable, so that $\opt = \omega\left(\frac{1}{\epsilon}\right)$.  The question we ask is: {\em how many pings are necessary and sufficient for obtaining such a guarantee?}

\smallskip \noindent {\bf Roadmap.} We will first focus on a very simple class of policies where each packet only pings on arrival (if at all). We term these \poa policies, as discussed earlier. 
Section~\ref{sec:poa} and Appendix~\ref{sec:ub2} respectively 
study  \poa policies when the departure is constant-rate and Poisson while the arrival is adversarial. Appendix~\ref{sec:unif_arrival} extends these results to the opposite case where the arrival process is constant-rate or Poisson. 
In Section~\ref{sec:lb_dep} we show that \poa policies are essentially optimal under these restrictions on departures or arrivals, while they are insufficient for general arrival and departure processes. For the latter setting, we present a continuous pinging policy \pico{} and analyze its performance in Section~\ref{sec:pico}.

\section{\poa Policy for the Constant-rate Departure Case}
\label{sec:poa}
\label{sec:benchmark}
We first consider the constant-rate departure setting, where the arrival process is completely arbitrary but the departure process generates one token per step.\footnote{This is without loss of generality by scaling the arrival and departure rates by the same factor. Further, the total number of packets generated by \poa remains unchanged regardless of the departure rate. As mentioned before, we only assume $\opt = \omega(1/ \eps)$ to keep the setting interesting.} Therefore, in each time step, if the queue is non-empty, one packet departs the system. An arbitrary number of packets can arrive at any time step; all these packets will have the same arrival time, but we will assume an arbitrary but fixed ordering among  these packets. For our results, we can view the arrivals as being determined by an adversary, albeit one who is oblivious to the randomness of the pinging policy (i.e., does not know which packets have pinged so far).

\subsection{Optimality of \poa Policies for Constant-rate Departures} 
We begin with the observation that for constant-rate departures, if packet $i$ sends a ping at time $t$ with its current height $h_{it}$, the server knows that its height at all times $t' \ge t$ is $\max\{h_{it} - (t' - t), 0\}$. In fact, the server also knows the packet's height at previous time steps, assuming the packet has entered the queue. Furthermore, since a packet can only look ahead into the queue, the information set of this packet (defined in Section~\ref{sec:model}) is the same at time $t = a_i$ (arrival time) and at all subsequent times. This immediately yields the following structural property.

\begin{lemma}
\label{lem:opt_dep}
For any pinging policy for constant-rate departures, an equivalent policy (in terms of the expected number of pings and the reconstruction error) has packets only sending pings on arrival. 
\end{lemma}

More formally, for any pinging algorithm $\cP$, we can construct an algorithm $\cP'$ in which when packet $i$ arrives, we compute the total probability that $\cP$ will send a ping (in the future; we can do this since the packet's information set at all times is known), and send a ping at arrival with that probability. The server will have at least the same amount of information as in the case of $\cP$.

The lemma implies that any policy can be characterized by a single function $f : \mathbb{N} \mapsto [0,1]$. If a packet arrives and finds the height of the queue is $h$, it sends a ping to the server with probability $f(h)$, else it never sends a ping. This shows that the optimal policy belongs to the class of \poa policies.

\subsection{The \poa Policy and its Analysis} 
We begin with a simple \poa policy and analyze its performance. The policy is as follows. 

\begin{itemize}
\item Ping algorithm $\cP$ (packet side; run independently for each packet): Let $h$ be the queue height after the packet arrives. Then it sends a ping with probability $f(h) := \min\left(1, \frac{2 \ln(1/\epsilon)}{\epsilon h} \right)$. The ping consists of the arrival time and the height $h$.
\item Estimation algorithm $\cE$ (server side; run at each time step): At time $t$, if the last packet received was at time $t'$ with height $h$, the server's estimate $e_t = \max(0, h - (t' - t))$. In other words, the server's estimation assumes there were no arrivals since the last ping.
\end{itemize}

In the rest of this section, we will show the following theorem. It turns out to achieve an essentially optimal number of pings; we show a matching lower bound in Section~\ref{sec:lb_dep}.

\begin{theorem}
\label{thm:unif_dep}
Consider the queue monitoring problem over a time interval $[0, T]$ where packet arrivals $\cA$ are arbitrary and departures are contant-rate at a unit rate. Assuming the queue heights at times $0$ and $T$ are zero, the \poa policy described above achieves the bound: $ \E[ \alg ] \le 37 \epsilon \cdot \opt.$
\end{theorem}

We begin the proof by setting up notation. Assume that the arriving packets $\cA$ are numbered $1, 2, \dots, N$, and that the arrival times $a_i$ are in non-decreasing order. (For convenience, even if packets arrive in the same slot, assume that the order is respected, as is natural in an application like traffic.) We denote the height of the queue just {\em after} the arrival of packet $i$ but before the arrival of packet $i+1$ by $h(a_i)$. As before, let $T$ denote the time horizon at which all packets have departed. 

Because of the unit departure rate assumption, the packet stays in the queue for exactly $h(a_i)$ steps. Therefore, the quantity $\opt$ (defined as the time average of the height) satisfies:
\begin{equation*}
    \opt = \frac{1}{T}\sum_{t=1}^{T} h_t = \frac{1}{T} \sum_{i \in \cA} h(a_i).
\end{equation*}

Since we have a unit departure rate, the update satisfies the property that $e_t \le h_t$ at all points of time. Thus the error term of interest is therefore
\begin{equation}
\label{eq:alg3}
    \alg = \frac{1}{T} \sum_{t=1}^T \E[(h_t - e_t)].
\end{equation}

\smallskip \noindent {\bf Accounting for {\sc \alg} via packet lags. }
\label{sec:lag}
Define the {\em lag} of packet $i$, denoted $\delta_i$, as the additional expected time it takes for the server to become ``aware'' of packet $i$. Formally, if packet $i$ arrives at time $t$ and $j$ is the first packet $\ge i$ to send a ping to the server, then $\delta_i$ would be $\min(h(a_i), a_j - a_i)$.\footnote{To be precise, the minimum must also include the term $(T - a_i)$ for handling the packets arriving in the last few time steps. This is a mild technicality that we will ignore here since $T$ is be assumed to be sufficiently large.} Note that if $i$ itself sends a ping, this ensures  $\delta_i = 0$.

With this notation, we now observe that a packet's contribution to the numerator of  $\alg$ (Eq~(\ref{eq:alg3})) is exactly equal to $\delta_i$.  This is because the quantity $(h_t - e_t)$ can be viewed as the number of packets that the server is unaware of at time $t$. Combining these formulas, we therefore have 
\begin{equation} 
\label{eq:alg_opt}
\frac{\alg}{\opt} = \frac{\sum_{i \in \cA} \delta_i}{\sum_{i \in \cA} h(a_i)}.
\end{equation}

Note that the denominator is deterministic and depends solely on the packet arrival and departures, while the numerator is an expectation that depends on the pinging algorithm. 
The main theorem is now the following, which when combined with Eq~(\ref{eq:alg_opt}) implies Theorem~\ref{thm:unif_dep}.

\begin{theorem}
\label{thm:ping}
The pinging strategy given by $f$ satisfies $ \sum_i \delta_i  \le 37 \epsilon \cdot \sum_i h(a_i)$.
\end{theorem}

\subsubsection{Proof of Theorem~\ref{thm:ping}}
The rest of this section is devoted to proving Theorem~\ref{thm:ping}. The idea of the proof is to consider the expression for $\delta_i$ and show that it is small if sufficiently many packets arrive ``soon after'' packet $i$. There can be packets for which this may not happen, and for these packets, we show how to charge the $\delta_i$ values to the sum of the heights (or the area under the height curve) in a certain time interval, which is the contribution of this interval to the delay $\opt$.

Let us focus on packet $i$, arriving at time $a_i$. We assume that $h(a_i) > \frac{2 \log (1/\epsilon)}{\epsilon}$, as otherwise the packet sends a ping on arrival with probability 1, and thus $\delta_i = 0$.  Let us now write an expression for $\delta_i$. Define $p(i,j)$ to be the probability that $(i+j)$ is the first packet $\ge i$ to send a ping.  Then by definition, we have
\begin{equation}\label{eq:delta-definition}
\delta_i = \sum_{j \ge 0} p(i, j) \min \{ (a_{i+j} - a_i), h(a_i) \}.  
\end{equation}

(The sum thus runs over all the packets in the interval $[1, T]$ that arrive after $i$.)  
Next, from the choice of our ping probability, the intuition is that one of the packets in the set $\{i, i+1, \dots, i+ \eps h(a_i)\}$ will send a ping. Let $S_i = \{0,1,\ldots, \eps h(a_i)\}$ capture the relevant indices.  
Now consider the prefix of $\delta_i$ as $\gamma_i = \sum_{j \in S_i} p(i, j) \min \{ (a_{i+j} - a_i), h(a_i) \}.$

\begin{lemma}
For any packet $i$, we have $\delta_i \le \gamma_i + \epsilon~  h(a_i)$.
\end{lemma}
\begin{proof}
Consider the packets $i + j$, for $j \in S_i$. For all these packets, we have that the ping probability (no matter when they arrive) is at least $\frac{2 \log(1/\epsilon)}{\epsilon (1+\epsilon) h} > \frac{ \log(1/\epsilon)}{\epsilon h}$, where $h = h(a_i)$. This is because $h(a_{i+j}) \le (1+\epsilon) h$ for these packets.  

Since each packet $i+j$ for $j \in S_i$ sends a ping with probability at least $\frac{ \log(1/\epsilon)}{\epsilon h}$, we have:
\[ \sum_{j > \epsilon h} p(i,j) \le \left(1 - \frac{ \log(1/\epsilon)}{\epsilon h}\right)^{\eps h}   \le \eps\]
Therefore $ \sum_{j > \epsilon h}p(i, j) \min \{ (a_{i+j} - a_i), h(a_i) \} \le \eps h(a_i)$. This completes the proof.
\end{proof}

Combining with Equation~(\ref{eq:alg_opt}), we have 
\begin{equation}
\label{eq:alg_opt2}
\frac{\alg}{\opt} = \frac{\sum_i \delta_i}{\sum_i h(a_i)} \le \frac{\sum_i \gamma_i}{\sum_i h(a_i)} + \epsilon.
\end{equation} 

Therefore, to show the theorem, we need to bound $\sum_i \gamma_i$. We do this as follows. We make a sweep over the packets. At every step, we maintain a ``leftover'' (or unaccounted) $\gamma$ mass, which is the sum of $\gamma_j$ values for a certain range of $j < i$. We then charge the sum of this and the $\gamma_j$ values for all packets that arrive in the interval $[a_i, t_i]$, where $t_i$ is an appropriately chosen time. For interval $[t_1,t_2]$, denote the area under the height curve (or total delay) as: $ A[t_1,t_2] = \sum_{t = t_1}^{t_2} h_t$. We will charge the total leftover $\gamma$ to $A[a_i, t_i]$. 

Formally, $U$ denotes the unaccounted $\gamma$ mass so far. Define the following:

\begin{define}
\label{def:ti}
\[t_i = \min\left\{\frac{h(a_i)}{3}, a_{i + \eps h(a_i)} - a_i\right\},\]
and let $r_i \in S_i$ denote the largest index s.t. $a_{i + r_i} \le a_i +  t_i$.
\end{define}

\begin{algorithm}[htbp]
 \caption{The {\sc Charging} Analysis}
   \label{alg:charge}
\begin{algorithmic}
    \STATE Set $i \leftarrow 0$; $U \leftarrow 0$; $k \leftarrow 0$.
    \WHILE {$i \le N$}
    \STATE Compute $t_i$ and $r_i$ as in Definition~\ref{def:ti}.
     \IF {  $\gamma_j \le 3 \gamma_i$ for all $j \in [i, i+r_i]$ } 
        \STATE $U \leftarrow U + \sum_{j=i}^{i+r_i} \gamma_j$; and $\tilde{A} \leftarrow  A[a_i, t_i]$.
        \STATE Charge $U$ to $\tilde{A}$. 
        \STATE Set $U \leftarrow 0$; and $i \leftarrow i + r_i + 1$.
        \ELSE 
        \STATE $\ell \leftarrow$ Smallest index with $\gamma_{\ell} > 3 \gamma_i$ and $\ell \le i + r_i$.
        \STATE $U \leftarrow U + \sum_{j=i}^{\ell-1} \gamma_j$; and $i \leftarrow \ell$.
     \ENDIF
     \ENDWHILE
\end{algorithmic}
\end{algorithm}

Our charging scheme is presented in Algorithm~\ref{alg:charge}. Here, $N$ is the total number of packets.  Our key lemma is the following:

\begin{lemma}
\label{lem:charging}
Whenever $U$ is charged to $\tilde{A}$ in the above scheme, we have $U \le 36 \epsilon \tilde{A}$.
\end{lemma}
\begin{proof}
Consider any iteration of the while loop of Algorithm~\ref{alg:charge}, and let $i$ be the index of the packet at the start of the loop. We will show that the following invariant is always maintained at the start of the loop:

 \smallskip \noindent {\bf Invariant.} $U \le 3 \epsilon h(a_i) \gamma_i$. 

Denote $h = h(a_i)$ and $r = r_i$. First observe that $t_i \ge \gamma_i/3$. To see this, denote $\tilde{t}_i = a_{i + \eps h} - a_i$. Note that $\gamma_i \le \min(h, \tilde{t}_i)$. If $\tilde{t}_i \le h/3$, then $t_i = \tilde{t}_i \ge \gamma_i$. Otherwise, $t_i = h/3 \ge 3 \gamma_i$.

We now consider the two cases dictated by the if statement in Algorithm~\ref{alg:charge}.

 \smallskip \noindent {\bf Case 1:} For all $j \in [i, i+r]$, $\gamma_j \le 3 \gamma_i$.  In this case, noting that $r \le \eps h$, we have
\[\sum_{j \in [i, i+r]} \gamma_j \le   r (3\gamma_i) \le 3\epsilon h \gamma_i.\]
Since the $U$ at the start of the loop is $3 \eps h \gamma_i$ by the invariant, the final $U$ is at most $6 \eps h \gamma_i$.  The area $\tilde{A} = A[a_i, t_i]$ is at least $\frac{1}{2} h t_i$. Since $\gamma_i \le 3 t_i$, the area is always at least $\frac{1}{6} h \gamma_i$. Therefore, $U \le 36 \eps \tilde{A}$. The invariant is trivially satisfied for the next iteration.

 \smallskip \noindent {\bf Case 2:} There exists  $\ell \in [i, i+r]$ such that $\gamma_\ell > 3\gamma_i$.

In this case, we only need to prove that our invariant is maintained when the index moves to $\ell$. To see this, note that the initial $U$ is at most $3 \epsilon h \gamma_i$ by the invariant. Therefore, final $U$ satisfies
\[ U \le 3 \epsilon h \gamma_i + \sum_{j \in [i, \ell)} \gamma_i \le 6\epsilon h \gamma_i,  \]
where we use the fact that $\ell -i \le r < \epsilon h$ and that $\ell$ is the first index with $\gamma_\ell > 3\gamma_i$. Thus the total unaccounted mass $U$ at index $\ell$ is less than $6 \epsilon h \gamma_i$. We argue that this $U \le 3 \epsilon \gamma_\ell h(a_{\ell})$. Using the fact that $\gamma_\ell > 3\gamma_i$, it suffices to prove that $h(a_{\ell}) > \frac{2}{3} h$. Since $a_{i+r} - a_i \le h/3$, there are at most $h/3$ departures during this period (by the constant-rate departure assumption), so that $h(a_{\ell}) \ge \frac{2}{3} h$. This proves the invariant holds at the next iteration, and completes the proof of the lemma.
\end{proof}

Since the $\tilde{A}$ in different iterations correspond to areas for disjoint time intervals and since the total $U$ is the same as $\sum_{i \in \cA} \gamma_i$, we have:  $\sum_{i \in \cA} \gamma_i \le 36 \epsilon \sum_t h_t.$ Since the RHS is equal to $\sum_i h(a_i)$, we combine with Eq~(\ref{eq:alg_opt2}) to complete the proof of Theorem~\ref{thm:ping}.

\section{Optimality Results and Lower Bounds for \poa Policies}
\label{sec:lb_dep}
We will now present almost matching lower bounds on the number of pings needed to achieve $\eps$ reconstruction error (Eq~(\ref{eq:eps})) for the constant-rate departure and constant-rate arrival settings, showing the optimality of the \poa class of policies. We will also show that \poa policies stop being optimal when we further generalize the arrival or departure processes, motivating the \pico class of policies considered in Section~\ref{sec:pico}. All proofs are deferred to Appendix~\ref{sec:lb-proofs}.

\smallskip \noindent {\bf Constant-rate Departures.} 
We first consider the setting from Section~\ref{sec:benchmark}. We know from Lemma~\ref{lem:opt_dep} that the optimal policy is a \poa policy. We have the following lemma whose bound almost matches the ping probability of the \poa policy we analyzed in Section~\ref{sec:poa}.

\begin{lemma}
\label{lem:lb}
\label{cor:lb}
For $\eps < \frac{1}{8}$, let $f$ be a ping probability function in a \poa policy that achieves $O(\eps)$ reconstruction error.
Then for any $h = \Omega\left(\frac{1}{\eps}\right)$, we must have  $\sum_{j = h}^{(1+8\eps)h} f(j) \ge \frac{1}{6}$. Therefore, for $\eps < \frac{1}{8}$ and any monotone $f$ that achieves  $\eps$ reconstruction error, we have  $ f(h) = \Omega\left(\min\left(1,\frac{1}{\eps h} \right)\right)$.
\end{lemma}

 \smallskip \noindent {\bf Constant-rate Arrivals.}
In this setting, the optimal policy need not be a \poa policy. Nevertheless, in Lemma~\ref{lem:fixh}, we show for certain types of canonical instances where the queue height is almost surely close to a given value $h$, no policy can do much better than the \poa policy we analyzed in Appendix~\ref{sec:unif_arrival}, that is, the expected number of pings cannot be improved by more than a constant factor for these instances. 

Given any height $h$, for any instance, define $N_h = \left\{i \ | h_{ia_i} \in \left[h (1- 8 \epsilon), h  \right] \right\}$ as the number of arrivals (or time steps) with queue size in $\left[h (1 - 8 \epsilon), h \right]$. We show a lower bound for the number of pings required by {\em any} pinging policy below. 

\begin{lemma}
\label{lem:fixh}
Fix any $\epsilon \in (0,1/16)$. For any given value $h = \Omega\left(\frac{1}{\epsilon}\right)$, there exist instances with $N_h = N - o(N)$ (where $N$ is the total number of arrivals), such that any pinging policy that achieves $O(\eps)$ reconstruction error sends at least $\frac{1}{48 \epsilon h}$ pings per packet on average. 
\end{lemma}

The above lemma shows that no policy (not just \poa policies) can improve on the expected number of pings when the queue height stays close to any value $h$. As a corollary, it shows that any \poa policy that achieves $\epsilon$ reconstruction error requires $f(h) = \Omega\left(\frac{1}{\epsilon h}\right)$ on the instances from Lemma~\ref{lem:fixh}. The last bound matches the upper bound of the \poa policy.

\smallskip \noindent {\bf Insufficiency of \poa for General Stochastic Arrivals or Departures.} 
As our final lower bound, we show that when we consider stochastic processes that are more bursty than Poisson, \poa policies are insufficient to achieve $\epsilon$-reconstruction error. This motivates the \pico class of policies in Section~\ref{sec:pico}.

\begin{lemma}
\label{lem:eg}
There exist instances with $i.i.d.$ arrival and departure processes and a constant $\epsilon > 0$ such that no \poa policy can achieve $\epsilon$-reconstruction error.
\end{lemma}

\section{General Arrivals and Departures: The \pico Policy}
\label{sec:pico}
In this section, we consider the most general case when both the arrivals and departures are arbitrary processes, either stochastic or adversarial. Lemma~\ref{lem:eg} shows that \poa policies are insufficient for this case, and we need a novel set of policies. Below, we present more nuanced examples that show that even pinging on both arrival and departures is insufficient. These examples make a case for policies that ping {\em continuously}, albeit at a rate that is continuously adjusted based on their current queue height and waiting time. 

 \smallskip \noindent {\bf Justification for Continuous Pinging Policies.}
 One might wonder about policies that ping on departure in addition to arrival. However, this can either lead to large reconstruction error or a large number of pings.  

\begin{example}
\label{eg1}
Initially, an unknown number $h$ of packets arrive into the system. Now  consider two scenarios: In the first scenario, all packets depart in one step, while in the latter scenario, all but one packets depart, and the remaining packet stays for $h$ steps.
\end{example}

 It is an easy exercise to show that the server cannot achieve $\eps$-reconstruction error without distinguishing the two scenarios. We will show that if packets ping only on arrival or departure, then all packets must ping on arrival. First note that only pings from the last packet can help the server with distinguishing the two scenarios. Further, a ping on departure is useless, since the server needs to maintain an estimate for the $h$ steps till this packet departs (in the second scenario), and will hence incur a large error by the time this ping is received. Thus, the last packet must ping on arrival if pings are made only on arrival or departure. Finally, it remains the case that a packet must ping even if it is not the last packet, as it does not see packets arriving after it. Therefore, we cannot achieve $\eps$-reconstruction error unless every packet pings on arrival. 

The next example goes a step further and makes a case for continuous pinging.

\begin{example}
\label{eg3}
Initially, an unknown number $h$ packets arrive into the system and the $i$-th arriving packet departs in either $2^i$ or $2^{i+1}$ time steps. 
\end{example}

Suppose the $i$-th packet is the last packet. Then, the server must know its existence because otherwise it doesn't know if the queue is empty or not for at least $2^i$ steps after the first $i-1$ packets depart. The resulting error of $2^i$ is unacceptable as the total time the first $i-1$ packets stay in the queue is also $\Theta(2^i)$. In this case the server can't learn about the $i$-th packet without a ping from it, and just as in the previous example, this ping cannot be on departure. Finally, it remains the case that it must ping even if it is not the last, as it doesn't see packets arriving after it. This motivates the need for each packet to ping continuously.

\subsection{The \pico Policy} 
We now present our {\em ping continuously} or \pico policy. Let $w_{it} := t - a_i$ denote packet $i$'s waiting time in the queue at time $t$. We define $h_{it}$ to be the number of packets ahead of packet $i$ at time $t$. We make the mild assumption that  
the minimum waiting time is $1$ unit (i.e., a packet cannot arrive and leave the same time instant).

\begin{itemize}
    \item Ping algorithm $\cP$ (packet side): At each time $t \in (a_i, d_i]$, packet $i$ sends a ping with probability 
    \begin{equation}
        \label{eqn:pico-prob}
     g(h_{it}, w_{it}) = \frac{5 \ln (1/\epsilon)}{\eps^2}\frac{1}{w_{it} h_{it}}.
     \end{equation}
    The ping consists of its waiting time $w_{it}$ and its current height $h_{it}$.
    \item Estimation algorithm $\cE$ (server side): At each time $t$, the server does the following step:
    
    \begin{itemize}
        \item Let $P_t$ denote the set of packets paired with the times when they sent a ping so far. That is, $(i, t') \in P_t$ means that packet $i$ sent a ping at time $t' \leq t$.
        \item For each $(i, t') \in P_t$ the server sets its estimated height $e_{tit'}$ as:
        \begin{equation*}
         e_{tit'} = \begin{cases}
            h_{it'} (1+3 \epsilon) &\text{if  $t \le t' + 3 \epsilon w_{it'}$}\\
            0 &\text{otherwise}
        \end{cases}
        \end{equation*}
        \item The server's estimate of height is $e_{t} = \max_{(i,t') \in P_t} e_{tit'}$.
    \end{itemize}
\end{itemize} 

Roughly speaking, if the server receives a ping from a packet with height $h$ and waiting time $w$, it guesses that the height is $\ge h$ for $3 \eps w$ steps in the future. If the queue suddenly empties, the error will not be too large as it can be `charged' to the previous waiting cost using the fact that the packet waited for $w$ steps and had a height $\ge h$ all that time. We also illustrate the policy via an example:
\begin{example}
Suppose $h$ packets arrive in the system at time $0$, and they all depart at time $1$. Suppose at time $0$, a subset of the packets sends a ping. Let $h'$ denote the largest height of any packet that sends a ping. Note that the waiting time is $1$. Then the server maintains an estimate of $h'(1+3 \epsilon)$ till time $1 + 3 \epsilon$, and subsequently sets the estimate to zero.
\end{example}

\newcommand{\cL}{\mathcal{L}}
\newcommand{\cU}{\mathcal{U}}

\subsection{Algorithmic Intuitions and Analysis Overview}
Before jumping into the analysis, we give more intuitions behind the policy and how they are naturally tied to the analysis. Recall that our goal is to approximately reconstruct the time-varying height profile  $\{h_t\}$, $t \in [T]$ from pings. We can visualize it by having a stack of $h_t$ points at time $t$ where $y$th point from the bottom represents the $y$th oldest one among packets alive at time $t$. Let $(t, y)$ denote the point and $j(t, y)$ the corresponding packet. For brevity let $i = j(t, y)$. We know that the packet $i$ has waited for $w_{it}$ steps since its arrival and it had at least $h_{it}$ packets (including itself) ahead of it all the time. It means that the rectangle of width $w_{it}$ and height $h_{it}$ having $(t, y)$ as its upper and right corner must lie inside the profile. Let $\cL_{it}$ denote the rectangle. 

Rectangles $\cL_{it}$ have several important properties. First, rectangle $\cL_{it}$ only depends on packet $j(t, y)$'s waiting time and height at time $t$. Thus, the packet can communicate the rectangle to the server without consulting with other packets and this is what a ping encodes conceptually. Second, the union of all rectangles $\cL_{it}$ is exactly equal to the time-varying height profile. 

\begin{figure}[htbp] 
    \centering
\includegraphics[width=0.9\linewidth]{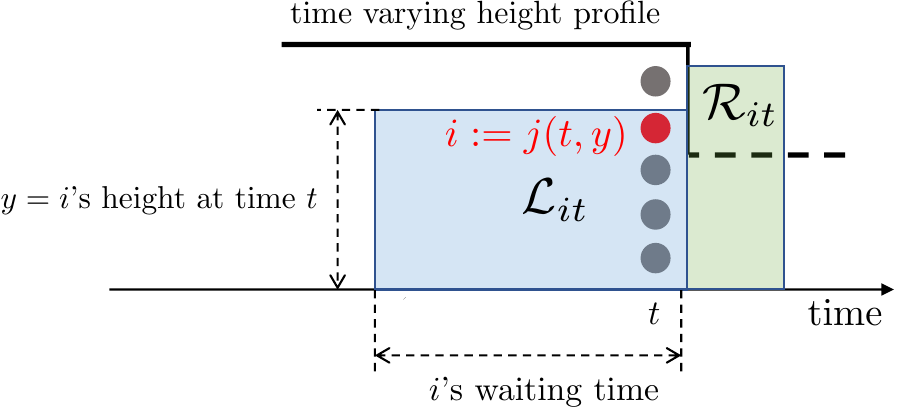}
  \caption{Visualization of \pico. Suppose the packet $i  = j(t, y)$ of height $y$ pings at current time $t$. The packet has had height at least $y$ since it arrival and therefore the rectangle $\cL_{it}$, colored blue, of height $h$ and width equal to $i$'s waiting time at time $t$ fits under the time varying height profile. The server projects $\cL_{it}$ into the future. The projection $\cR_{it}$, colored green, is a vertical and horizontal extension of $\cL_{it}$. The server's estimation is the union of the extended rectangles.}
  \label{fig:pico-alg}
\end{figure}

But, there are are some issues. Communicating one rectangle means one ping. Thus, to ping parsimoniously, a rectangle is communicated with probability inversely proportional to its area. Another issue is that each rectangle $\cL_{it}$ only contains past information. Thus, the server projects the rectangle into the future by stretching it by a small factor. The crux of the analysis lies in showing the height profile is approximately sandwiched by the union of communicated rectangles $\cL_{it}$ and the union of their future projections. See Figure~\ref{fig:pico-alg} for visualization of this discussion.

\subsection{Analysis}

\newcommand{\area}{\textsc{Area}}

Our main result about the \pico policy is the following.
\begin{theorem}
\label{thm:pico}
Consider the queue monitoring problem over a time interval $[0, T]$ where packet arrivals and departures are arbitrary. Assuming the queue heights at the initial and final time steps are zero, for any $\epsilon \in (0,1/5)$, the \pico policy described above achieves the bound $ \E[ \alg ] \le 10 \epsilon \cdot \opt$.
\end{theorem}

Recall the definition of $\alg$ from Eq~(\ref{eq:algt}).
Now consider the following geometric view of the pinging process. The $x$-axis represents time, and the $y$-axis represents queue height. If the height of the queue at integer time $t$ is $h_t$, any point $(t',y)$ for $y \le h_t$ and $t' \in (t-1,t]$ is marked as full. The full points -- denote them $\cH$ -- in this $2$-D space capture the height profile of the queue, and we call this the {\em height diagram}. We let $j(t, y)$ denote the packet corresponding to point $(t, y) \in \cH$, which is the packet that has height $y$ at time $t$.

Now, when the server gets a ping from packet $j(t, y)$ at time $t$ with height $y$ and waiting time $w(t, y) := w_{j(t, y), t}$, it creates a rectangle $\cR_{ty} := [t, t + 3\epsilon \cdot w(t, y)] \times [0, (1+3\eps) y]$. 
For brevity, we may say $(t, y)$ has waiting time $w(t, y)$ and height $y$, instead of saying $j(t,y)$ does. Similarly, we may say $(t, y)$ pings when $j(p, y)$ does. Let 
\[ \cR_t = \bigcup_{y, t' \le t} \cR_{t'y} \]
denote the union of rectangles the server has created until time $t$. The server uses the height of $\cR_t$ at the current time $t$ as the queue height estimate $e_t$ at the  time. Note that since any rectangle constructed at time $t$ only covers current and future time steps, $\cR_t$ is not affected by pings, arrivals, or departures at future time steps.

The analysis consists of two parts, bounding the under-estimation error and the over-estimation error. Formally we will show $\E [ \area ( \cH \setminus \cR_T)] \leq \eps \area (\cH)$ (Corollary~\ref{cor:under-error}) and $\area ( \cR_T \setminus \cH) \leq 9\eps \area (\cH)$ (Lemma~\ref{lem:upper-error}), which will complete the proof of Theorem~\ref{thm:pico}. Here, $\area(\cdot)$ means the area of what is inside in the parentheses; thus, the total delay is  $\area(\cH)$.

\subsubsection{Bounding Under-estimation Error}

We first show the under-estimation error is low. The following key lemma claims that our estimation covers any point in height diagram with large probability. 

\begin{lemma}
    \label{lem:pico-under-key}
    Consider any point $(t, y)$ in $\mathcal{H}$. Then, $\Pr[ (t,y) \notin \cR_t] \le \epsilon$.
\end{lemma}

\begin{corollary}
    \label{cor:under-error}
        $\E \left[\area(\cH \setminus \cR_T) \right] \leq \eps \area(\cH)$.
\end{corollary}

  The following lemma articulates our high-level proof strategy to show Lemma~\ref{lem:pico-under-key}.

\begin{lemma}
    \label{lem:pico-A}
    For any $(t, y) \in \cH$, there exists a set $A$ of points in $\cH$ such that i) at least one point $(t', y') \in A$ pings with probability $1 - \eps$ and ii) $(t, y) \in \cR_{t'y'}$ for all $(t', y') \in A$.
\end{lemma}

In words, if a point in $A$ pings, it adds a rectangle to our estimation that covers $(t, y)$ and at least one point in $A$ pings with large probability. It is easy to see Lemma~\ref{lem:pico-A} immediately implies 
Lemma~\ref{lem:pico-under-key}, and therefore, Corollary~\ref{cor:under-error}. Thus, the remainder of the section is devoted to proving Lemma~\ref{lem:pico-A}. We fix a point $(t, h) \in \cH$ throughout the section.

Finding a desirable $A$ is not straightforward. Note that i) and ii) are somewhat in conflict. To cover more points along with the fixed point $(t, y)$ by rectangles created in the past, intuitively they should be large. But as discussed before, the rectangles are created with probability inversely proportional to their respective area. Thus, if we want to use large rectangles to cover the point, we will have to use more to guarantee a large success probability. Note that the obvious approach of setting $A := \{ (t', y') \; | \; t' \in [t - \eps w(t, y)], y' \in [y - \eps y, y]\}$ doesn't work because some points in $A$ may create too large rectangles (particularly if they have a large waiting time) and therefore they are  created with too small probabilities.

To find a proper $A$, we find a special point for each  $y' \in [(1 - \eps)y, y]$. 

\begin{lemma}
    \label{lem:pico-special-point}
    There exists a point $q_{y'} := (\tau_{y'}, y')$ for each $y' \in [(1 - \eps)y, y]$ such that every point $(t', y')$ with  $t' \in [\tau_{y'} - \eps w_{y'}, \tau_{y'}]$ has waiting time between $(1 - \eps) w_{y'}$ and $4 w_{y'}$, and $t - \tau_{y'} \leq \frac{1}{3} \eps w_{y'}$, 
    where $w_{y'} := w(\tau_{y'}, y')$ denotes the waiting time of $q_{y'}$.
\end{lemma}
\begin{proof}
Fix a $y' \in [(1 - \eps)y, y]$. Let $t_1 := t$. To find a desired $q_{y'}$, we will iteratively find a sequence of $p_1 := (t_1 := t, y'), p_2 := (t_2, y'), ..., p_{K-1} := (t_{K-1}, y'), q_{y'} :=  p_K := (t_K := \tau_{y'}, y')$. 
Let $w_k := w(t_k, y')$. Let $p_{k+1}$ be any point in $\{ (t', y') \; | \; t'  \in [t_k -\eps w_k, t_k] \}$ such that $w_{k+1} > 4 w_k$, if any; if there's no such point, then we set $K = k$. Clearly, this process must terminate for fixed $t$, since $w_k$ increases exponentially in $k$ and $w_k \leq t$. Note that $w_k < \frac{1}{4} w_{k+1} < \frac{1}{4^2} w_{k+2} < \ldots < \frac{1}{4^{K - k}} w_K$.

By definition of $p_K$, any point $(t', y')$ with $t' \in [t_K - \eps w_K, t_K]$ has waiting time at most $4 w_K$. Further, the packet $j(t_K, y')$ must have height at least $y'$ before time $t_K$. This implies the packet $j(t', y')$ arrives no later than $j(t_K, y')$. Thus, $(t', y')$ must have waiting time at least $w_K  - \eps w_K$ as $t' \in [t_K - \eps t_K, t_K]$. Knowing $w_{y'} = w_{K}$ by definition, we have the first claim. 

The second claim follows from an easy algebra:
\begin{align*}
    t - \tau_{y'} 
    &=     t_1 - t_K \\
    &=    \sum_{k = 1}^{K-1} (t_{k} - t_{k+1}) \\ 
    &\leq \eps \sum_{k = 1}^{K-1}  w_k \\ 
    &\leq 
    \eps \sum_{k = 1}^{K-1} \frac{1}{4^{K - k}} w_K \\
    %
    &\leq \frac{1}{3} \eps w_K = \frac{1}{3} \eps w_{y'}.
\end{align*}
\end{proof}

Intuitively, $q_{y'}$ is a point that creates a rectangle that covers $(t, y)$ if it pings. Further, all points $\{(t', y') \; | \; t' \in [\tau_{y'} - \eps w_{y'}, \tau_{y'}]\}$ have similar waiting times. Thus, they create rectangles of similar areas that cover $(t, y)$. We include those points in $A$.

\begin{lemma}
    Let $A_{y'} := \{(t', y') \; | \; t' \in [\tau_{y'} - \eps w_{y'}, \tau_{y'}]\}$ and $A := \bigcup_{y' \in [(1 - \eps)y, y]} A_{y'}$. Then, $A$ satisfies the properties stated in Lemma~\ref{lem:pico-A}.
\end{lemma}
\begin{proof}
    Recall that each point $(t', y') \in A_{y'}$ has waiting time at most $4 w_{y'}$ (Lemma~
        \ref{lem:pico-special-point}).
    and therefore pings with probability at least $\frac{5 \ln( 1/ \eps)}{\eps^2} \frac{1}{4 w_{y'} y'}$ (Eqn.~(\ref{eqn:pico-prob})). Thus, the probability that no point in $A$ pings is at most, 
    \begin{align*}
   & \prod_{y' \in [(1 - \eps)y, y]} \prod_{t' \in  [\tau_{y'} - \eps w_{y'}, \tau_{y'}]} (1 - \frac{5 \ln( 1/ \eps)}{\eps^2} \frac{1}{4 w_{y'} y'}) \\
    &\leq \prod_{y' \in [(1 - \eps)y, y]} \prod_{t' \in  [\tau_{y'} - \eps w_{y'}, \tau_{y'}]} \exp( - \frac{5 \ln( 1/ \eps)}{\eps^2} \frac{1}{4 w_{y'} y'})\\
    &\leq \prod_{y' \in [(1 - \eps)y, y]}  \exp( -\frac{1.25 \ln ( 1/ \eps)}{\eps} \frac{1}{y})  \\
    &= \prod_{y' \in [(1 - \eps)y, y]}  \exp( -\frac{1.25 \ln ( 1/ \eps)}{\eps} \frac{\eps y}{y})  \leq
    \eps, 
    \end{align*}
    which shows property i).

    To show property ii), consider any $(t', y') \in A$. Our goal is to  show that $(t', y')$ creates a rectangle covering $(t, y)$ if it pings, i.e.,  $(t, y) \in \cR_{t'y'}$.
    Note $\cR_{t'y'} = [t', t' + 3\eps w(t', y')] \times [0, (1+3\eps) y']$. Clearly we have $t' \leq t$.
    \begin{align*}
        & \;\;\; t' + 3\eps w(t', y') - t \\
        &\geq \tau_{y'}  - \eps w_{y'} + 3\eps w(t', y') - t  &\mbox{[Definition of $A$]}\\
        &\geq \tau_{y'}  - \eps w_{y'} + 3\eps ( 1- \eps) w_{y'} - t  &\mbox{[Lemma~\ref{lem:pico-special-point}]}\\
        &\geq t - \frac{1}{3}\eps w_{y'}  - \eps w_{y'} + 3\eps ( 1- \eps) w_{y'} - t &\mbox{[Lemma~\ref{lem:pico-special-point}]}\\
        &\geq 0 &\mbox{[When $\eps \in (0, 5/9]$]}\\
    \end{align*}
    Thus, we also have $t \leq t' + 3\eps w(t', y')$. Finally, since $y' \in [(1 - \eps)y, y]$, we have $0 \leq y \leq (1+3\eps) y'$. Therefore, we have shown $(t, y) \in \cR_{t'y'}$.
\end{proof}

\subsubsection{Bounding Over-estimation Error}

We now switch to bounding the over-estimation error. Any $(t,h) \in \cH$ corresponds to a packet $i$ that exists in the queue at time $t$. We therefore overload terminology and denote this as $(i,t) \in \cH$. To measure the over-estimation error of $\cH$ by $\cR_T$, we view $\cH$ as the union of $\cL_{it} := [t  - w_{it}, t] \times [0,  h_{it}]$ for $(i, t) \in \cH$. The proof of the next lemma immediately follows from the observation that the height of the packet corresponding to $(i, t)$ can only decrease over time.  

\begin{lemma}
    \label{lem:lb-set}
    For every $(i, t) \in \cH$, we have $\cL_{it} \subseteq \cH$. Therefore, we have $\bigcup_{(i, t) \in \cH} \cL_{it}  \subseteq \cR_T $.
\end{lemma}

Now for the sake of analysis, we consider the union of 
$\cU_{it} := [t  - w_{it}, t + 3 \eps w_{it}] \times [0,  (1+ 3 \eps)h_{it}]$, which is an extension of $\cL_{it}$. Note that $\cU_{it}$ fully contains both $\cL_{it}$ and $\cR_{it}$. Therefore, their union is a super-set of $\cR_T$, which implies the next lemma.

\begin{lemma}
    \label{lem:ub-set}
 $\cR_T \subseteq \bigcup_{(i, t) \in \cH} \cU_{it}$.
\end{lemma}

We finally bound the over-estimation error by bounding the area of the regions $\cU$.

\begin{lemma}
    \label{lem:upper-error}
    For any $\eps \in (0, 1/3), \area(\cR_T \setminus \cH) \leq 9 \eps \area(\cH)$.
\end{lemma}
\begin{proof}
    From Lemmas~\ref{lem:lb-set} and \ref{lem:ub-set}, it suffices to show that 
    \[\area(\bigcup_{(i, t) \in \cH} \cU_{it}) \leq (1 + 3\eps)^2 \area(\bigcup_{(i, t) \in \cH} \cL_{it}).\]
    Let $\cU'_{it} := [t  - w_{it}, t + 3 \eps w_{it}] \times [0,  h_{it}]$. Since $\cU_{it}$ stretches $\cU'_{it}$ vertically by a $(1+3\eps)$ factor keeping the rectangle abutting the $x$-axis, we clearly have 
    $\area(\bigcup_{(i, t) \in \cH} \cU_{it})    = (1 + 3\eps) \area(\bigcup_{(i, t) \in \cH} \cU'_{it})$.
    
    Thus, we only need to show that $\area(\bigcup_{(i, t) \in \cH} \cU'_{it}) \leq (1 + 3\eps) \area(\bigcup_{(i, t) \in \cH} \cL_{it})$. Note that $\cU'_{it}$ is equivalent to a rectangle that results from extending $\cL_{it}$ by $(1+3\eps)$ factor to the right. The claim then follows by considering each height $h' \in [0, h]$: if we stretch the interval at height $h'$ in each $\cL_{it}$ by a uniform factor, and the length of their union stretches by the same factor or less. 
\end{proof}

As discussed, this lemma, combined with Corollary~\ref{cor:under-error}, proves Theorem~\ref{thm:pico}.

\subsection{Comparison with \poa Policies} 
We finally consider how much worse is the number of pings of this policy compared to the \poa policy. We show that the the number of pings is $\tilde{O}\left(\frac{\log h}{\epsilon}\right)$ factor larger for constant-rate arrivals or departures.

\begin{lemma}
\label{lem:approx_length}
For constant-rate arrivals or departures, for packets arriving when the queue height is $h$, the number of pings of the \pico policy is within  $O\left(\frac{ \ln(1/\epsilon)}{\epsilon} \cdot \ln h \right)$ of  the \poa policy.
\end{lemma}
\begin{proof}
First consider the constant-rate departure case. Here, a packet arriving at height $h$ sends a ping after $t$ steps with probability $\frac{5 \ln (1/\epsilon)}{\epsilon^2} \frac{1}{t(h-t+1)}$. Therefore, the total number of pings sent by this packet (in expectation) is
\begin{equation}
    \label{eq:count}
\sum_{t=0}^h \frac{5 \ln (1/\epsilon)}{\epsilon^2} \frac{1}{t(h-t+1)} = O\left( \frac{\ln (1/\epsilon)}{\epsilon^2} \frac{\ln h}{h} \right)
\end{equation} 
This is a factor $O\left(\frac{\ln h}{\epsilon} \right)$ worse than the \poa policy that would have pinged with probability $O\left( \frac{\ln(1/\epsilon)}{\epsilon h} \right)$ for this packet. Similarly, for the constant-rate arrival case, suppose the current height of the queue is $h$. Then a packet with height $j$ arrived at least $h-j$ steps ago, so that its waiting time is at least $h-j+1$. Therefore, the expected number of pings at this step is again at most the quantity from Eq~(\ref{eq:count}).
Therefore, the overhead is again a factor of $O\left(\frac{\ln h \ln(1/\epsilon)}{\epsilon} \right)$ more than the \poa policy, which would have pinged with probability $O\left( \frac{1}{\epsilon h} \right)$ at this step. 
\end{proof}

\section{Conclusion}

In this paper, we presented the queue monitoring problem as an abstraction for tracking congestion in traffic applications. We showed decentralized online algorithms that achieve near-optimal trade-offs between the number of pings and the error in queue length. Our work shows that the question of congestion monitoring is non-trivial even for a single queue.

The most important direction for future work is to combine congestion monitoring with algorithms for optimal routing in a network to minimize delay based on this congestion information, much like low-complexity methods for load balancing~\cite{JIQ,scalable,AL1,AL2,SankarL04} that have been widely studied in networking and queueing theory. Such algorithms do not follow in any obvious way from our current work, and we leave addressing this as an interesting open question.

Another interesting direction is the following: the \poa policy provides an unbiased estimator and is therefore likely robust. In contrast, the \pico policy is not unbiased as it is, and it would be very interesting to develop another policy that is unbiased. Finally, it would be interesting to study the effect of noise in our problem.


\newpage
\appendix
\section{\poa Policy for the Poisson Departure Case}

\label{sec:poisson}
\label{sec:ub2}

In this section we design and analyze a \poa policy when departures follow a Poisson process instead of being constant-rate. This generalizes a $G/M/1$ queue, and allows for variation in inter-departure times. 
The advantage of our guarantee for Poisson departures is that it directly yields the same guarantee per queue for a {\em network} of G/M/1 queues (generalizing a {\em Jackson network}~\cite{grimmett2001}) when each queue runs the above pinging policy, since in such a network, all departure processes are Poisson.

We will assume time is continuous, and will consider the quantities $\opt$ and $\alg$ in the limit as the time horizon goes to infinity. Following the notation in Section~\ref{sec:model}, we have:

\begin{equation}
    \label{eq:optt2}
\opt  =\lim_{T \rightarrow \infty} \frac{1}{T} \E \left[ \sum_{i \in S} (d_i - a_i) \right]  = \lim_{T \rightarrow \infty} \frac{1}{T} \E\left[ \int_{t=0}^T h_t dt \right]
\end{equation}

\begin{equation}
    \label{eq:algt2}
\alg = \lim_{T \rightarrow \infty}\frac{1}{T} \E\left[\int_{t=0}^T | h_t - e_t | \right]
\end{equation}

We will need the technical assumptions that the inter-arrival times and independent of the service times, and that the arrival process causes the queue to be ergodic. The latter is a milder and considerably more general assumption than $i.i.d.$ inter-arrival times, though slightly more restrictive than completely adversarial. For instance, the sequence of arrivals could be dictated by distributions that depend on states of a hidden Markov chain. 

\subsection{Algorithm}
We will build on the analysis in Section~\ref{sec:benchmark} to address the case where the time between departure tokens is distributed $i.i.d.$ as {\tt Exponential}$(\mu)$, independent of arrival times. We assume an arbitrary arrival process, subject only to the mild assumption that the resulting queueing system is ergodic. 
The ping policy is given below: 

\begin{itemize}
    \item Ping algorithm $\cP$ (packet side):  When a packet arrives at time $t$, with queue height of $h$ after arrival, it sends a ping with probability $f(h) = \min\left(1, 2\frac{ \log(1/\epsilon)}{\epsilon h}\right)$. The ping consists of the pair $(t,h)$. 
    \item Estimation algorithm $\cE$ (server side): Immediately after receiving a ping $(t, h)$, the server sets $e_t = h$. To maintain the height estimate, since the server does not know the exact departure times of the packets, it updates $e_t$ as follows: It generates a sequence of ticks such that the time between ticks are $i.i.d.$ {\tt Exponential}$(\mu)$.\footnote{Though we could have updated $e_t$ uniformly with time, updating via an Exponential clock is simpler to analyze.} On each tick, if $e_t > 0$, it is decreased by $1$.
\end{itemize}

\subsection{Analysis}
At time $t$, let $i_t$ denote the last packet to have sent a ping, and let $c_t$ denote its current height. This is unknown to the server, since the exact departure times are unknown. Nevertheless, we can use triangle inequality to write $\alg$ from Eq~(\ref{eq:algt}) as:
\begin{equation} \label{eq:algsum}
\alg  \le  Q + R
\end{equation} 
where
\begin{align*}
Q &= \lim_{T \rightarrow \infty} \frac{1}{T} \E\left[\int_{t=0}^T (h_t - c_t) dt \right]; \mbox{ and } \\
R &= \lim_{T \rightarrow \infty}  \frac{1}{T} \E\left[\int_{t=0}^T |e_t - c_t | dt \right] 
\end{align*}
We call $Q$ the arrival error and  $R$ the departure error. We will bound these two quantities separately.

\medskip \noindent {\bf Bounding the Arrival Error.}
We will first use the property that for any ergodic system, the time average of any statistic is the same as the average of the statistic seen by a Poisson process. This is the so-called PASTA property~\cite{grimmett2001,Wolff82}. Note that the departure process is as follows: A sequence of tokens arrives as a Poisson process with mean time between two tokens being $1/\mu$. When the token arrives and the queue is non-empty, one packet departs the system. 

Denote this sequence of tokens by $1,2,\ldots, r, \ldots,N$, and let $h_r$ denote the queue height just before token $r$ arrives; similarly $c_r$ is the height of the last pinged packet just before this token arrives. Note that for large $T$, the number of tokens in $T$ steps is $N = \mu T$ almost surely. Using the PASTA property, we therefore have:
\begin{align*}
 \opt &= \lim_{N \rightarrow \infty} \E\left[\frac{1}{N} \sum_{r=1}^N h_r \right]; 
\mbox{ and }  \\
Q &= \lim_{N \rightarrow \infty} \E\left[ \frac{1}{N} \sum_{r=1}^N (h_r - c_r) \right]
\end{align*}
where the expectation is over the randomness in the number of arrivals between two tokens, and the pings.

Fix the randomness in the arrival process, {\em i.e.}, the realization of the tokens. Then the above quantities are the same as the delay and error respectively had we replaced ``tokens" with ``time units", where one packet deterministically departs the system each time unit. The analysis in Section~\ref{sec:benchmark} combined with taking expectations over the arrival process yields
\begin{equation}
\label{eq:q1}
Q \le 37 \epsilon \cdot \opt
\end{equation}

\medskip \noindent {\bf Bounding the Departure Error.}
Consider the Poisson process that combines the departure tokens of rate $\mu$, and the server's ticks of rate $\mu$ that are used to update $e_t$. This combined Poisson process has rate $2 \mu$. Denote the ticks in the new process as $1,2,\ldots, s, \ldots, 2N$. 

Denote the height of the queue and its estimate just before the arrival of the $s^{\text{th}}$ tick as $h_s$ and $e_s$ respectively. At each tick $s$, independently of previous ticks, with probability $1/2$, $h_s$ decreases by $1$ if $h_s > 0$, and with the remaining probability, $e_s$ decreases by $1$ if $e_s > 0$.  Using the PASTA property, we can write
\begin{align*}
 \opt &= \lim_{N \rightarrow \infty} \E\left[\frac{1}{2N} \sum_{s=1}^N h_s \right]; \mbox{ and} \\
R &= \lim_{N \rightarrow \infty} \E\left[ \frac{1}{2N} \sum_{s=1}^N |e_s - c_s| \right]
\end{align*}

We split this summation based on the identity of the packet whose current height corresponds to $c_s$. Let $R_i$ denote the expected contribution from arrival $i$, ignoring the $\frac{1}{N}$ term in the denominator.

Consider some arrival $i$ that contributes to the above summation. Suppose the first contribution is at tick $s_0$, and suppose the height of the queue just after this arrival's arrival was $h$. We will condition on this event. This packet contributes to the above summation only if it sent a ping.  

To upper bound its contribution, condition on this packet sending a ping. Since subsequent arrivals lead to new pings that truncate the contribution of the ping for $i$, for an upper bound, we assume no subsequent arrivals to $i$. Then the quantity $\delta_s = e_s - c_s$ starts at $\delta_{s_0} = 0$, and for subsequent ticks $s \ge s_0$ performs a random walk that increases by $1$ with probability $1/2$ and decreases by $1$ with probability $1/2$. In $2h$ steps, either $e_s$ or $c_s$ or both have become zero, therefore, for every subsequent tick, $|\delta_s|$ decreases by $1$ with probability $1/2$ and stays the same with probability $1/2$. Note finally that $|\delta_s| \le h$ always.

By Azuma's inequality, at $s_1 = 2h + s_0$, we have $ \Pr[ |\delta_{s_1}| > \alpha] \le 2 e^{- \alpha^2/4h}$. Setting $\alpha = 4\sqrt{h \ln h}$, we have
$ \Pr[|\delta_{s_1}| > 4 \sqrt{h \ln h}] \le \frac{2}{h^4}$.

In the event $|\delta_{s_1}| = h_1 > 4 \sqrt{h \ln h}$, this quantity decreases by $1$ in expectation every two ticks. Therefore, conditioned on this event we have $ \E\left[\sum_{s= s_1}^{\infty} |\delta_s|\right] \le 2 h_1 \le 2h$. Note further that for $s_0 \le s \le s_0 + 2h$, the probability that $|\delta_s| > 4 \sqrt{h \ln h}$ at any step is at most $\frac{4}{h^3}$ by union bounds.

Therefore, we can bound the contribution to the expected error conditioned on this arrival sending a ping as:
\begin{align*} 
 \frac{1}{2} \E\left[\sum_{s=s_0}^{\infty} |\delta_s|\right]   &\le  4h \left(\sqrt{h \ln h}\right) + \frac{4}{h^3} \left(2h^2 +2h\right) \\
  &\le 4 \sqrt{h^3 \ln h} + \frac{32}{h}  \le 8 \sqrt{h^3 \ln h}
 \end{align*}

Since the packet pings with probability $f(h) \le 2 \frac{\ln(1/\epsilon)}{\epsilon h}$, the expected contribution from this arrival, conditioned on it being at height $h$ is:
\begin{equation*} 
\E[R_i]  \le 2\frac{\ln(1/\epsilon)}{\epsilon h} \cdot 8 \sqrt{h^3 \ln h} 
 \le 16 \frac{\ln(1/\epsilon)}{\epsilon} \sqrt{h \ln h} 
 \end{equation*}
For $h = \Omega\left(\frac{1}{\epsilon^5}\right)$, the above quantity is bounded by $4 \epsilon h$. Therefore $ \E[R_i] \le 4 \epsilon h + O\left(\frac{1}{\epsilon^5}\right)$.

Since this packet $i$ contributes $h$ to the expected total delay, summing these contributions, removing the conditioning on the arrival time of this packet, and dividing by $N$, we have 
\begin{equation}
\label{eq:R1}
R \le 4 \epsilon \cdot \opt + O\left(\frac{1}{\epsilon^5}\right)
\end{equation}

Combining Equations~(\ref{eq:algsum}),~(\ref{eq:q1}), and~(\ref{eq:R1}), we have the following theorem.

\begin{theorem}
When departure tokens arrive with inter-arrival time distributed as $i.i.d.$ {\tt Exponential}$(\mu)$ and the system is ergodic, for any $\epsilon \in (0,1)$, there is a pinging policy with:
\[ \alg \le  O(\epsilon) \cdot \opt + O\left(\frac{1}{\epsilon^5}\right). \]
\end{theorem}

\section{\poa Policy for the Constant-rate Arrivals}
\label{sec:unif_arrival}
In this section, we  show that a \poa policy is also effective in the case where {\em arrivals} are constant-rate and departure tokens are generated by an arbitrary (potentially adversarial) process.  

Our policy is simply \poa with ping probability $f(h) = \min\left(1, \frac{4}{\epsilon h} \right)$, and we show that it achieves $\alg \le 4 \epsilon \cdot \opt$. 
%
There is one key difference from the constant-rate departure case in Appendix~\ref{sec:ub2}. Since the departure process is adversarial, in our policy, the server keeps its estimate  constant between pings. This introduces further error in the estimate; however, the constant-rate arrival process ensures this does not accumulate.  Formally, the \poa policy is the following.

\begin{itemize}
    \item Ping algorithm $\cP$ (packet side): When packet $i$ arrives at time $t$ and the queue height is $h$ just after, it pings with probability $\min\left(1, \frac{4}{\epsilon h} \right)$; the ping consists of the pair $(t, h)$.
    \item Estimation algorithm $\cE$ (server side): Upon receiving a ping $(t, h)$, the server sets $e_t = h$ (in case of multiple pings at the same time, the largest $h$ is kept). In between pings, the server keeps $e_t$ constant.
\end{itemize}

We are interested in the average error in the server's estimation of the queue height (recall the definition of $\alg$ from Eq~(\ref{eq:algt})). We will show the following theorem in Appendix~\ref{sec:ub3}.

\begin{theorem} 
\label{thm:mg1}
Suppose packets arrive at unit rate in a time interval $[0, T]$, and suppose the departure process is arbitrary. If the queue heights at $0$ and $T$ are equal, then we have
\[ \alg = \frac{1}{T} \E \left[ \sum_t |h_t - e_t| \right] \le 4\eps \cdot \opt . \]
\end{theorem}

\smallskip\noindent {\bf Poisson Arrivals.} In Appendix~\ref{sec:ub4}, we derive \poa policies for Poisson arrivals and adversarial departures, and derive the analog of Theorem~\ref{thm:mg1} for this setting.

\section{\poa Policy for the Constant-rate Arrivals: Proof of Theorem~\ref{thm:mg1}}
\label{sec:ub3}
Recall the time-average delay, $\opt$, from Eq~(\ref{eq:optt}). Recall that we denote arrival times by $a_1, a_2, \ldots$ and departure times by $d_1, d_2, \ldots$. Let $\cA$ denote the set of arrival times, and $\cD$ denote the set of departure times. Let $h(a_i)$ denote the queue size just before the arrival of packet $i$, and $h(d_i)$ denotes the queue size just after the departure of packet $i$.   

In order to analyze our pinging policy, it will help to view the arrival and departure times as being ordered (packet $i$ arrives before packet $j$ if $i< j$).\footnote{We make this assumption even if the packets arrive/depart in the same time interval (breaking ties if needed).}

Let $\cA$ be the set of arrivals in the interval and $\cD$ be the set of departures. Then for any $\eps \in (0, 1/2)$, we will first show that the policy satisfies 
\begin{equation}
    \label{eq:unif_arr}
 \E \left[ \sum_t  |h_t - e_t| \right] \le 2 \epsilon \left( \sum_{i \in \cA} h(a_i) + \sum_{i \in \cD} h(d_i) \right).
\end{equation}

The main idea behind our analysis is to view the term $\sum_t |h_t - e_t|$ as a sum of contributions from arriving and departing packets. Consider the execution of the algorithm, and suppose that at time $t$, the last ping was received at $t' \le t$. Then $(h_t - e_t)$ is precisely equal to $A_t - D_t$, where $A_t$ (resp., $D_t$) is the number of arrivals (resp., departures) since the last ping. Thus we can upper bound $|h_t - e_t|$ by $A_t + D_t$. This means that it suffices to upper bound $\sum_t \E[ A_t ]$ and $\sum_t \E[D_t]$.

Let us first consider $\sum_t A_t$. Similar to Section~\ref{sec:benchmark}, for any arriving packet $i$, its contribution to the sum is exactly equal to the number of time steps between $a_i$ and the next ping. Let us denote this quantity by $\delta_i$. In other words, we have $\sum_t A_t = \sum_{i \in \cA} \delta_i$. 

A key technical lemma here is the following. 

\begin{lemma}\label{lem:arrival-delta-lemma}
For any arrival packet $i$, our pinging policy satisfies
$\E[ \delta_i ] \le 2 \epsilon h(a_i)$.
\end{lemma}
\begin{proof}
Let us consider any arrival packet $i \in \cA$, and for convenience, denote $h = h(a_i)$. Now if $h \le \frac{4}{\epsilon}$, the ping probability of packet $i$ is 1, and thus $\delta_i = 0$.  Thus we may assume that $h > \frac{4}{\epsilon}$.

We will use the expression $\E[\delta_i] = \sum_{r \ge 1} \Pr [ \delta_i \ge r]$ and bound the terms in the summation separately.\footnote{To be precise, we must truncate this sum when the arrival time exceeds $T$, thus the expression above is formally an upper bound.}

We observe that for any $r \ge 1$, we have 
\[ \Pr[ \delta_i \ge r] = \prod_{j = 0}^{r-1} \left( 1- p(i,j) \right), \]
where $p(i,j)$ is the ping probability for the $(i+j)$th packet. For this quantity to be maximized, we must have $p(i,j)$ as small as possible, which means that the height of the $(i+j)$th arrival must be as large as possible. As exactly one packet arrives at each step, $(h+j)$ is an upper bound on the height. This implies that
\[
\Pr[\delta_i \ge r] \le \prod_{j=0}^{r-1} \left( 1 - \frac{1}{\eps (h+j)} \right) 
\le e^{ -\sum_{j=0}^{r-1} \frac{1}{\eps(h+j)}}
\]
Noting that $\sum_{j=0}^{r-1} \frac{1}{h+j} \ge \int_0^r \frac{1}{h+x} dx$, we have:
\[
\Pr[\delta_i \ge r]  \le e^{- \frac{1}{\eps} \log \left( \frac{h+r}{h} \right)} = \left( 1+\frac{r}{h} \right)^{-1/\eps}. 
\]
Thus, putting everything together, we have that
\[ \E[ \delta_i ] \le \sum_{r\ge 1} \left( 1+\frac{r}{h} \right)^{-1/\eps}. \]
Writing $k = 1/\eps$ for convenience, we have 
\[ \sum_{r\ge 1} \left( 1+\frac{r}{h} \right)^{-k} \le \int_{0}^{\infty} \left(1+\frac{x}{h} \right)^{-k} dx = \frac{h}{(k-1)}. \]
The last equality follows by making the change of variable $y= x/h$ and performing a standard computation. Noting that $\eps < 1/2$, the lemma follows.
\end{proof}

Next, consider the sum $\sum_t D_t$. Once again, for the $i^{th}$ departure (which occurs at time $d_i$), its contribution to the sum is exactly equal to the number of time steps between $d_i$ and the {\em next} ping. This is because a packet $i$ contributes to $D_t$ if it has departed at $d_i \le t$, and there has been no ping in $[d_i,t]$.  Let us denote this quantity by $\delta_i'$. For instance, if the packet that arrives at time $d_i+1$ sends a ping, then $\delta_i' =1$. Thus we have $\sum_t D_t = \sum_{i \in [N]} \delta_i'$. We show that:

\begin{lemma}\label{lem:departure-delta-lemma}
For any departure $i$, our pinging policy satisfies $\E[ \delta_i'] \le 2 \epsilon h(d_i)$.
\end{lemma}
\begin{proof}
The proof is almost identical to that of Lemma~\ref{lem:arrival-delta-lemma}. We still outline it for completeness. Consider any departure $i \in \cD$, and suppose that it happens when the queue has height $h = h(d_i)$. The key point is that for any $j \ge 1$, the $j$th arrival packet following departure $i$ has an arrival height $\le h +(j-1)$, and thus we can use exactly the computation above to conclude that $\E[ \delta_i' ] \le 2 \eps h(d_i)$, proving the lemma.
\end{proof}

Putting the lemmas together with our observations above,
\[ \E [\sum_t |h_t - e_t| ] \le 2 \eps \left[ \sum_{i \in \cA} h(a_i) + \sum_{i \in \cD} h(d_i) \right].\]

This completes the proof of Equation~(\ref{eq:unif_arr}). To complete the proof of Theorem~\ref{thm:mg1}, we first observe that because we have constant-rate arrivals, $\sum_{i \in \cA} h(a_i) = \sum_{t = 1}^{T} h_t = T \cdot \opt$. 

Next, if the queue heights at the beginning and end are the same, then we argue that
\[ \sum_{i \in \cA} h(a_i) = \sum_{i \in \cD} h(d_i). \]  
This can be seen as follows: suppose we view the change in height of the queue as a particle moving along the line graph (one vertex for every value of the height). Then if the particle starts and ends at the same node, we have that for any height $\ell$, the number of transitions from $\ell$ to $(\ell +1)$ is the same as the number of transitions from $(\ell+1)$ to $\ell$.  The first kind of transitions correspond to terms $h(a_i)$, while the second correspond to terms  $h(d_i)$. Thus the number of times $\ell$ appears in the two summations is exactly the same, for all $\ell$. This completes the proof of Theorem~\ref{thm:mg1}. 


\section{Generalization to Poisson Arrivals}
\label{sec:ub4}
We now consider the more general setting where arrivals follow a Poisson process instead of being constant-rate.  The Poisson arrival case yields a generalization of the classic M/G/1 queue. 

We assume that time is continuous. Packet arrivals are distributed as a Poisson process with rate $\lambda$ independent of the departure process. The time between departure tokens follows some distribution $Z$ such that $\lambda \E[Z] < 1$. We do not assume the the tokens are generated $i.i.d.$; all we need is that the resulting system is ergodic, so that statistics like the limiting distribution of queue heights and time-average delay are well-defined. 

Recall the time-average delay, $\opt$, from Eq~(\ref{eq:optt2}). From Appendix~\ref{sec:unif_arrival}, recall that we denote (random) arrival times by $a_1, a_2, \ldots$ and departure times by $d_1, d_2, \ldots$. Let $\cA$ denote the set of arrival times, and $\cD$ denote the set of departure times. Let $h(a_i)$ denote the queue size just before the arrival of packet $i$, and $h(d_i)$ denotes the queue size just after the departure of packet $i$.   

The pinging process is exactly the same as in Appendix~\ref{sec:unif_arrival}.  Recall the definition of $\alg$ from Eq~(\ref{eq:algt2}). We will build on the result in Appendix~\ref{sec:unif_arrival} to show the following theorem:

\begin{theorem} 
\label{thm:mg11}
When the arrival process is Poisson, and the system is ergodic, for $\epsilon \in (0,1/2)$, there is a  policy with:
\[\alg \le 4 \epsilon \cdot \opt. \]
\end{theorem}

Let $N = |\cA| = |\cD|$. Let $e(a_i)$ denote $e_t$  just before the arrival of packet $i$. The PASTA property combined with Eq~(\ref{eq:algt2}) implies
\begin{equation} 
\label{eq:alg2}
\alg = \lim_{N \rightarrow \infty} \frac{1}{N} \E\left[ \sum_{i \in \cA}  |h(a_i) - e(a_i)| \right]
\end{equation}

Fixing a sequence of arrivals, from the perspective of arriving packets, this system is exactly identical to that in Section~\ref{sec:unif_arrival} if we replace ``arriving packet" with ``time step". In other words, there is one arrival per time step, and arbitrarily many departures. Therefore, Equation~\ref{eq:unif_arr} holds as is, and yields:
\begin{align}
    \label{eq:poisson1}
 \lim_{N \rightarrow \infty} \frac{1}{N} \E \left[ \sum_{i \in \cA} |h(a_i) - e(a_i)| \right]  \\
 \le  2 \epsilon \lim_{N \rightarrow \infty} \frac{1}{N} \left( \sum_{i \in \cA} h(a_i) + \sum_{i \in \cD} h(d_i) \right) \nonumber
\end{align} 

For any ergodic system, in steady state, for any value $x$, $\Pr_i[h(a_i) = x] = \Pr_i[h(d_i) = x]$. This is because the system transitions with the same frequency from a state with $x$ packets in the queue to $x+1$ packets in the queue as the other way around. When combined with the PASTA property that the time-average queue height is the same the average queue height seen by arrivals, this implies the time-average queue height is the same as the average queue heights just after departures. Therefore, Eq~(\ref{eq:optt2}) implies the following for Poisson arrivals.
\begin{equation}
    \label{eq:opt2}
\opt = \lim_{N \rightarrow \infty} \frac{1}{N} \E\left[\sum_{i \in \cD} h(d_i) \right] = \lim_{N \rightarrow \infty} \frac{1}{N} \E\left[\sum_{i \in \cA} h(a_i)\right]
\end{equation} 

Combining  Equations~(\ref{eq:alg2}),~(\ref{eq:poisson1}), and~(\ref{eq:opt2}), this will show Theorem~\ref{thm:mg11}.



\section{Missing Proofs from Section~\ref{sec:lb_dep}}
\label{sec:lb-proofs}

\begin{proof}{of Lemma~\ref{lem:lb}.}
Suppose there exists an $h$ for which the desired inequality fails to hold. The instance has multiple phases $1, 2, \ldots, L, L+1$, where phase $i$ has the maximum queue length of $H_i + h$, where $H_i$ is sampled from $I:= [0, 8 \eps h]$ independently of other phases. At the very beginning, $h$ packets arrive instantly and the queue length stays in $[h, (1 + 8 \eps) h]$ throughout all phases (except the final phase). At the beginning of phase $i$, the queue length is $h$ and $H_i$ new packets arrive to increase the queue length to $h + H_i$. Subsequently no packets arrive so the queue length reduces as one packet departs each time step. The phase stops when the queue length becomes $h$, and the next phase begins. Note that phase $i$ lasts for $H_i$ time steps. The last phase is special in that no new packets arrive. In the analysis we will ignore the last phase for simplicity as it will have a vanishing effect when $L$ is sufficiently large. 

To make the argument clear, we assume that the server is notified when a new phase starts. This is w.l.o.g. as it can only strengthen the server. However, if the server receives no ping in phase $i \neq L+1$, which occurs with probability at least $5 / 6$ regardless of the value of $H_i$, the server only knows that the initial queue length  is in $[h, (1 + 8 \eps) h]$. It is an easy exercise to show that $t$ time steps after the phase $i$ starts, the expected error incurred by the server is $\max \{8 \eps h - t, 0\} / 2$, regardless of the server's estimate at the time. We know that no ping is made in phase $i$ and $H_i \geq 4\eps h$ with probability at least $5 / 12$. Therefore, in this phase the expected error is at least $2\eps h  \cdot 4 \eps h$, which follows from considering the first $4\eps h$ times steps of the phase. Thus, the total error is at least $10 \eps^2 h^2 / 3$ in expectation. In contrast the total queue height in a phase is at most $( 1+ 8 \eps) h \cdot 8 \eps h$. Therefore, the reconstruction error is at least $\frac{10\eps^2 h^2 / 3}{ ( 1+ 8 \eps) h \cdot  8 \eps h} = \Omega(\eps)$. 
\end{proof}

\begin{proof}{of Lemma~\ref{lem:fixh}.}
    As in the proof of Lemma~\ref{lem:lb}, the instance has multiple phases $1, 2, \ldots, L, L+1$, where phase $i$ has the minimum queue length of $h - H_i$, where $H_i$ is sampled from $I:= [0, 8 \eps h]$ independently of other phases. In the first phase, $h$ packets arrive and subsequently, $H_1$ packets depart. In each phase $i \ge 2$, one packet arrives per step till the queue length becomes $h$; subsequently $H_i$ packets depart and the queue length becomes $h - H_i$. Therefore the queue length stays in $[(1 - 8 \eps) h, h]$ throughout all phases except phase $1$. The server is notified when a new phase starts, but does not know $H_i$. If the probability of pinging is at most $\frac{1}{48 \epsilon h}$ on average,  the server receives no ping in phase $i$ in at least $5 / 6$ of the phases regardless of the value of $H_i$. In such a phase, the server only knows that the initial queue length is in $[(1 - 8 \eps) h, h]$. Repeating the argument in the previous proof, this implies the reconstruction error is $\Omega(\eps)$ when $L$ is large.
\end{proof}

\begin{proof}{of Lemma~\ref{lem:eg}.}
We consider the following instance. At each step, independently of previous time steps, $h$ packets arrive with probability $1/3$. Similarly, independently of previous time steps, if there are at least $h$ packets in the queue, then $h$ packets depart with probability $2/3$.

Since the queue height is always non-negative and departure occurs with a higher probability than arrival, it is easy to see that the height is $0$ or $h$ at a constant fraction of times over a long time period and $\opt = \Theta(h)$. We give the server more power and even assume it knows each time when the height changes from $0$ to $h$. At the next time step, the height could become $2h$, $0$, or remain unchanged. The class of \poa policies do not ping unless the height changes to $2h$. Thus, the server cannot distinguish between $h$ and $0$, each of which occurs with a constant probability. Consequently, there exists a constant $\epsilon>0$ such that no \poa policy can have $\epsilon$-reconstruction error.
\end{proof}

\end{document}